\documentclass[prl,superscriptaddress,twocolumn,showpacs]{revtex4}
\usepackage{graphicx} 
\usepackage{amsmath}
\usepackage{amsfonts}

\begin{document}

\title{Interferometric and noise signatures of Majorana fermion edge states
in transport experiments}
\author{Gr\'egory Str\"ubi}
\affiliation{Department of Physics, University of Basel, 
CH-4056 Basel, Switzerland}
\author{Wolfgang Belzig}
\affiliation{Fachbereich Physik, Universit\"at Konstanz, 
D-78457 Konstanz, Germany}
\author{Mahn-Soo Choi}
\affiliation{Department of Physics, Korea University, Seoul 136-713, Korea}
\author{C. Bruder}
\affiliation{Department of Physics, University of Basel, 
CH-4056 Basel, Switzerland}

\pacs{71.10.Pm, 73.23.-b, 74.45.+c}


\begin{abstract}
Domain walls between superconducting and magnetic regions placed on
top of a topological insulator support transport channels for Majorana
fermions. We propose to study noise correlations in a
Hanbury Brown-Twiss type interferometer and find three signatures of
the Majorana nature of the channels. First, the average charge current in the outgoing leads vanishes.  Furthermore, we predict an anomalously large shot noise in the output ports for a vanishing average current
signal. Adding a quantum point contact to the setup, we
find a surprising absence of partition noise which can be traced back
to the Majorana nature of the carriers.
\end{abstract}
\maketitle

Elementary excitations (often called quasiparticles) of
condensed-matter systems can show features that are not displayed by
the bare particles that they are composed of. A striking example are
quasiparticles that show neither fermionic nor bosonic statistics but
an intermediate `anyonic' form~\cite{anyons}. Majorana fermions
appearing at the core of half-vortices in p-wave superconductors have
been predicted to exhibit anyonic statistics~\cite{ivanov}.
Theoretical proposals to observe their existence in tunneling
experiments~\cite{bolech2007, tewari2008} were devised.

Recently, the possibility to realize Majorana-like quasiparticles on
the surface of a three-dimensional topological insulator has attracted
a lot of attention (see Ref.~\onlinecite{majorana} and references
therein).  It has been shown that the domain wall of two superconducting
regions support transport channels for Majorana fermions~\cite{Fu2008},
and the interface of superconducting and magnetic regions give rise to
transport channels for chiral Majorana fermions~\cite{Akhmerov2009,Fu2009}.

Up to now, these new excitations have not been observed
experimentally, but a number of schemes to detect them has been put
forward.  These include interferometric structures in which electrons
are converted to Majorana fermions and 
back~\cite{Akhmerov2009,Fu2009,chung2011}, 
as well as scanning probe devices coupled to Majorana
edge states that detect resonant Andreev reflection~\cite{law2009}.
Also the measurement of the back-action of Majorana edge states to 
a coupled flux qubit could provide a hint of their existence~\cite{hou2011}.

In the proposals using interferometry~\cite{Akhmerov2009,Fu2009}, the
authors considered a two-terminal Mach-Zehnder setup. A magnetic
domain wall carrying chiral electronic excitations meets a
superconducting region, where the incoming electron channel is split
into two chiral Majorana fermion channels surrounding the
superconductor. At the opposite side of the superconductor, the
Majorana channels recombine to form an outgoing chiral electron
channel.  Depending on the phase change $2\pi\phi/\phi_0$ between the
two Majorana arms that is determined by their geometric length and the
number of vortices threading the superconductor, an incoming electron
is converted either to an outgoing electron or an outgoing hole. The
\emph{effective} flux $\phi$ threading the Mach-Zehnder interferometer
includes the actual magnetic flux due to vortices, as well as the
dynamical phase of the Majorana fermions; $\phi_0=h/e$ is the flux
quantum. The conductance $G_{12}$, where 1(2) stands for the
incoming(outgoing) lead, is periodic in $\phi/\phi_0$:
$G_{12}=(e^2/h)\cos(2\pi\phi/\phi_0)$, at zero bias and low
temperatures.  Negative conductances correspond to outgoing holes:
charge conservation is ensured because the superconductor is grounded,
\textit{i.e.}, this Mach-Zehnder interferometer is actually a
three-terminal device.  This form of the conductance shows the same
periodicity as a normal (non-superconducting) interference experiment.
Hence, there is a need for further signatures of Majorana physics
beyond the Mach-Zehnder setup.

The structure we have in mind is a Hanbury Brown-Twiss (HBT) type
interferometer built on the surface of a topological insulator. This
setup is inspired by recent proposals
\cite{Fu2009,Akhmerov2009,bose2010,chung2011} and is related
to the two-particle Aharonov-Bohm effect~\cite{Samuelsson2004}. 
We calculate the current cross-correlations in the
two outgoing leads of this interferometer and predict the possibility
to switch between negative and positive current cross-correlations by
tuning the magnetic flux threading the superconductor. 
Positive cross-correlations are remarkable since non-interacting
fermions will always show a negative sign \cite{buettiker:92}; see,
however, \cite{martin1996,boerlin2002,samuelsson:02}. 
The cross-correlations are predicted to be temperature-independent in a
reasonable range of temperature and at low voltages. As in
\cite{chung2011} we find that the cross-correlations vanish when only
one source is active as the consequence of the transport through
Majorana modes.

We then consider a setup that contains an additional quantum point
contact (QPC), similarly as in \cite{Fu2009}. Strikingly, the partition noise
associated to the quantum point contact is predicted to vanish, which
is an evidence of the neutrality, or equivalently, the Majorana nature,
of the charge carriers.

\begin{figure}
\centering
  \includegraphics[width=0.4\textwidth]{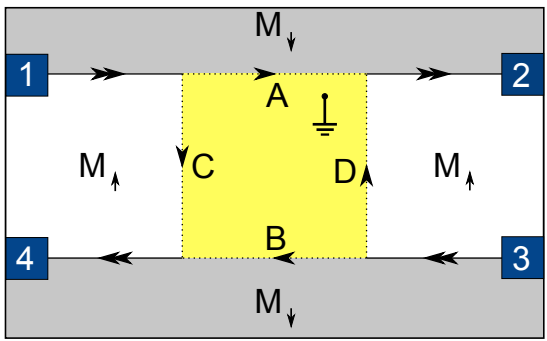}
  \caption{Hanbury Brown-Twiss type interferometer. A grounded
    superconductor surrounded by four magnetic domains is placed on
    the surface of a three-dimensional topological insulator.  The
    terminals 1, 2, 3, and 4 are connected to the outside circuit and
    biased at potentials $V_1, \ldots, V_4$ respectively. The
    magnetizations are chosen such that Dirac electron states at leads
    1, 2, 3, and 4 exist and propagate in the direction of the double
    arrows.  Electrons and holes can enter the interferometer at leads
    1 and 3, Majorana fermions propagate along the arms A,B,C, and D
    in the direction of the single arrows and electrons and holes
    leave through leads 2 and 4. A magnetic flux in the form of $n_v$
    vortices threading the superconductor will control the phase
    difference between the arms of the interferometer.  }
  \label{fig:HBT}
\end{figure}

We propose to realize a Hanbury Brown-Twiss type interferometer
consisting of a grounded superconductor surrounded by four magnetic
domains, as shown in Fig.~\ref{fig:HBT}.
In the outer arms 1, 2, 3, and 4, Dirac electrons propagate, while in
the center arms A, B, C, and D, at the edge of the superconductor,
only Majorana chiral fermions exist at energies below the induced
superconducting gap $\Delta$. The incoming Dirac electrons are
transformed into Majorana fermions which partially circle the
superconductor and are
converted back to Dirac electrons. This conversion process has already
been studied in \cite{Akhmerov2009,Fu2009} and is expected to be
perfectly symmetric: the Dirac electron is coherently split into the
two arms with equal probabilities. For the reversed process, a single
Majorana fermion is converted into a superposition of an electron and
a hole with equal probabilities as well.  An incoming electron can thus leave
as a hole, in which case a Cooper pair will flow into the grounded
superconductor. 
Figure~\ref{fig:HBT} does not show the backflow currents between
terminals 4 and 1 (2 and 3): since they are noiseless, they will not
affect any of our conclusions below. 

In the following, we would like to study the conductance and noise
properties of the interferometer when varying the magnetic flux,
\textit{i.e.}, the number of vortices threading the superconductor.

The Landauer-B\"uttiker formalism provides a straightforward analysis
of this interferometer once we know its scattering
matrix~\cite{buettiker:92}.  The scattering properties of the Dirac to
Majorana converter were established and discussed
in~\cite{Akhmerov2009,Fu2009}.  At zero energy, the full scattering
matrix is fixed by particle-hole symmetry. For small enough energies
$E \ll (v_M/v_F)\Delta$ where $v_M$ is the Majorana fermion velocity
and $v_F$ the electronic Fermi velocity at the surface of the bare
topological insulator, the following still holds
\begin{equation}
\left(\begin{array}{c}
a_2\\b_2\\a_4\\b_4
\end{array}\right) = 
\frac{1}{2}\left(\begin{array}{cccc}
1 & 1 & 1 & 1\\ 
1 & 1 & -1 & -1\\ 
1 & -1 & -\eta & \eta\\ 
1 & -1 & \eta & -\eta
\end{array}\right) 
\left(\begin{array}{c}
a_1\\b_1\\a_3\\b_3
\end{array}\right)\, .
\label{eq:scattering_matrix}
\end{equation}
The interferometric phase factor $\eta= e^{i 2\pi \phi/ \phi_0}
= (-1)^{n_v} e^{i E\delta L/\hbar v_M}$ has been concentrated to the
arm B by a gauge choice.  Here, $n_v$ is the number of vortices
threading the superconductor, $\delta L = L_A+L_B-L_C-L_D$ where $L_i$
is the length of arm $i$ of the interferometer. The operator $a_i$ ($b_i$)
is the annihilation operator of a Dirac electron (hole) in lead
$i$. The scattering matrix shown in Eq.~(\ref{eq:scattering_matrix})
is similar to the one obtained in \cite{chung2011}, which, however,
did not consider the possibility of vortices.

For topological reasons, one-particle quantities are not sensitive to
the enclosed flux in this structure: because of the chiral nature of
the Majorana states, no one-particle state will enclose the flux.  One
incoming electron or hole is scattered with equal probability to a
hole or an electron at lead 2 or 4. The outgoing currents thus vanish
on average. This vanishing conductance is a first hallmark of Majorana
fermions: in a standard setup with Andreev processes this could occur
only accidentally, and small perturbations would give rise to a
non-zero conductance. This vanishing conductance could in principle be
due to an interrupted circuit and has to be complemented by an
additional measurement of e.g. the current auto-correlation discussed
below.

On the other hand, when both sources are active we expect to see
a manifestation of an interesting two-particle Aharonov-Bohm 
effect~\cite{Samuelsson2004} for Majorana fermions.
As an example consider two incoming electrons in leads 1 and 3
 \begin{align}
a_1^\dagger a_3^\dagger = -(a_2^\dagger b_2^\dagger +\eta
a_4^\dagger b_4^\dagger)/2 - (\eta +1) (a_2^\dagger a_4^\dagger
-b_2^\dagger b_4^\dagger)/4 \nonumber\\ +(-\eta +1) (b_2^\dagger
a_4^\dagger- a_2^\dagger b_4^\dagger)/4\, .
\label{two-particle-inf}
\end{align}
The current cross-correlations between leads 2 and 4 are thus expected
to be sensitive to the parity of the number $n_v$ of enclosed vortices
through the phase parameter $\eta$. In particular, as shown later, it
is possible to switch between positive and negative cross-correlations
by tuning the magnetic field threading the superconductor. As a side
remark, note that post-selecting events with one fermion per lead
for $\eta =\pm 1$ yields maximally entangled pairs in particle-hole
space, and we can equivalently speak of anti-bunching~\cite{bose2010}.

Assuming the reservoirs connected to the incoming leads are specified
by the electron and hole distribution functions $n_{i,
  e}(E)=(\exp((E-eV_i)/k_BT)+1)^{-1}=1-n_{i,h}(-E)$ for lead $i$, we
can give explicit expressions for the current-current
correlations. The current cross-correlation between leads 2 and 4
$S_{24} = \frac{1}{2} \langle \{I_2, I_4\}\rangle$ 
are of special
interest:
\begin{equation}
S_{24} = -\frac{e^2}{h} \int_0^\infty dE\ \Re{(\eta(E))} (n_{1,e} -
n_{1,h})(n_{3,e} - n_{3,h})\, ,
\label{eq:hbt-cross-corr}
\end{equation}
which is sensitive to the magnetic flux through the real part of
$\eta$, $\Re{(\eta)} = (-1)^{n_v} \cos E\delta L/\hbar v_M$.  At
equilibrium $S_{24} = 0$, \textit{i.e.}, there is no thermal noise in
this quantity (electrons and holes compensate each other). This
temperature independence is expected to hold as long as
$k_{\mathrm{B}} T \ll  (v_M/v_F)\Delta$. For voltages $V_2 = V_4 = 0$, $V_{1} =
V_{3} = V$ (with respect to the potential of the superconductor),
temperatures such that $k_{\mathrm{B}} T\ll eV$ and an approximately
symmetric interferometer, $\delta L \ll \hbar v_M/eV$,
\begin{equation}
S_{24} = (-1)^{n_v+1}\frac{e^2}{h} \int_0^\infty dE\ (n_e + n_h) =
(-1)^{n_v+1}\frac{e^2}{h} e |V|\, .
\end{equation}
Thus, the sign of the cross-correlation is given by the parity of the
number of vortices. The possibility to achieve positive
cross-correlations for fermions is attributed here to electron-hole
conversions. 

We now look at the current auto-correlations in the outgoing
leads. While the outgoing current is zero on average, it is
carried by electrons and (the same number of) holes. Current
fluctuations are thus expected to be relevant. Indeed,
\begin{align}
 S_{22} = \frac{e^2}{h}\int_0^\infty
 dE\ [n_{1,e}+n_{1,h}+n_{3,e}+n_{3,h} \nonumber\\
-(n_{1,e}+n_{1,h})(n_{3,e}+n_{3,h})] \, .
\label{eq:hbt-auto-corr}
\end{align}
At zero bias, this reduces to the usual Johnson-Nyquist noise $S_{22}
= \frac{4e^2}{h} k_{\mathrm{B}}T$, while for voltages $V_2 = V_4 = 0$,
$V_{1} = V_{3} = V$ and $k_{\mathrm{B}}T\ll eV$, we obtain the shot
noise result $S_{22} = \frac{e^3}{h}|V|$, which is four times larger
than the maximal expected shot noise due to a beam splitter of chiral
electrons.  This remarkable result can be explained by noting that in
each scattering event both outgoing electrons and holes contribute to
the charge fluctuations, while giving a zero average current as a
consequence of the perfect electron-hole symmetry imposed by the
Majorana conversion.

\begin{figure}
\centering
  \includegraphics[width=0.4\textwidth]{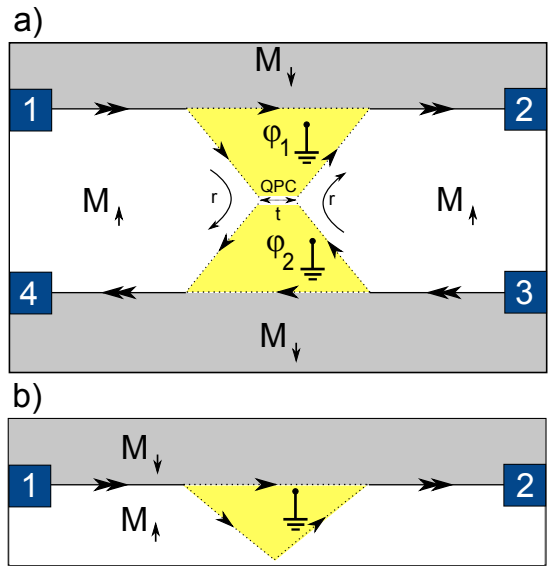}
  \caption{(a) Modified Hanbury Brown-Twiss interferometer. 
 Majorana excitations will propagate along the boundaries of the two
    triangular superconducting structures with phases $\varphi_1$,
    $\varphi_2$. An additional short gapped channel appears at
    the domain wall between the two superconducting regions, forming a
    quantum point contact characterized by reflection and transmission
    amplitudes $r$, $t$. 
The setup is similar to the one proposed in
    Ref.~\onlinecite{Fu2009}. 
 (b) In the fully transmissive case, $t=1$, $r=0$, the device splits into two
    separate Mach-Zehnder interferometers, one of which is shown here. 
}
  \label{fig:HBTQPC}
\end{figure}

We would now like to discuss a second possibility to obtain a
signature of Majorana fermions by adding a QPC to the previous setup,
see Fig.~\ref{fig:HBTQPC}a.  A novel feature will appear in the noise
properties, which we want to study in the same spirit as in the
previous section.

As explained in \cite{Fu2009}, the transmission and reflection 
amplitudes $t$, $r$ of the QPC can be strongly tuned by altering the
geometry of the QPC itself, or by changing the phase difference
$\varphi$ between the two superconducting parts.  A narrow
constriction would be dominated by direct tunneling and thus hardly
sensitive to the phase difference.  Therefore, the geometry we want to
consider is closer to a line junction supporting a non-chiral Majorana
channel on its own. By changing $\varphi=\varphi_1-\varphi_2$ from
$\varphi=0$ to $\varphi=\pi$, the channel appearing at the interface
of the two superconductors can be tuned from closed ($t\ll 1$) to
fully open ($t\lesssim 1$) at zero energy. For intermediate values of
the phase, the channel is gapped and the transmission amplitude
strongly depends on energy.

We would first like to look at the limiting cases. For $t=1$, $r=0$, the
upper and lower channels are not connected by the QPC. As a
consequence, the setup effectively reduces to two independent copies
of a Mach-Zehnder interferometer between terminals 1 and 2 (3 and 4)
(see Fig.~\ref{fig:HBTQPC}b).  The full current-current correlation
matrix $S_{\text{MZ+MZ}}$ for the outgoing leads is easy to obtain in
that case: the cross-correlations vanish since they are not connected
in any way, and the auto-correlations are given in Table \ref{tab:1}.
For $t=0$, the setup is equivalent to the HBT interferometer of the
previous section, whose correlation matrix $S_{\text{HBT}}$ is given
by Eqs.~(\ref{eq:hbt-cross-corr}), (\ref{eq:hbt-auto-corr}).  At
intermediate values of $t$, we use the same formalism as for the HBT
setup, taking the QPC into account in the scattering matrix:
\begin{equation}
\left(\begin{array}{c}
a_2\\b_2\\a_4\\b_4
\end{array}\right) = 
\frac{1}{2}\left(\begin{array}{cccc}
\eta_1 - t & \eta_1+t & r & r\\ 
\eta_1+t & \eta_1-t & -r & -r\\ 
r & -r & -\eta_2+t & \eta_2+t\\ 
r & -r & \eta_2+t & -\eta_2+t
\end{array}\right) 
\left(\begin{array}{c}
a_1\\b_1\\a_3\\b_3
\end{array}\right).
\end{equation}
Here, $\eta_{1(2)}$ is the interferometric phase factor for
Majorana fermions around the upper (lower) superconductor.

In this case, the average currents do not identically vanish. In fact
the conductances $G_{12} = G_{34} = \frac{e^2}{h}|t|$ are proportional
to the transmission \textit{amplitude}. This allows to experimentally
access the QPC properties. The two remaining conductances $G_{14}$ and
$G_{32}$ still vanish.

We now focus on the quantities of interest, namely the current-current
correlations $S_{22}$, $S_{44}$, and $S_{24}$. Proceeding through the
usual steps, the resulting $2 \times 2$ correlation matrix, for the
two outgoing leads, at fixed energy can be decomposed as
\begin{equation}
S = R\times S_{\text{HBT}} + T\times S_{\text{MZ+MZ}}\:,
\label{eq:cross-corr-qpc}
\end{equation}
where $R=|r|^2$ and $T=|t|^2$ are the reflection and transmission
probabilities of the QPC. The QPC effectively interpolates between the
two limiting cases: surprisingly, there are no mixed terms
proportional to $RT$; in other words, while there are the (auto and
cross-correlation) noise terms related to the HBT and MZ
interferometer present in the structure, there is no partition
noise. This is one of the main results of our paper and is deeply
rooted in the Majorana nature of the excitations transported along the
boundaries of the superconductor.

In the following we give an intuitive explanation of this
remarkable feature of Eq.~(\ref{eq:cross-corr-qpc}). Partition noise
in the context of an electronic beam splitter is due to the transport
of charge in discrete units. An incoming electron is coherently split
into e.g. two channels, and in a current measurement the electron will
contribute to the current in one, and only one, outgoing channel.  The
splitting thereby induces current fluctuations proportional to the
charge of the electron.  Majorana fermions, on the other hand, fail to
generate electric current fluctuations since they are neutral. We thus
believe that the absence of electronic partition noise predicted by
Eq.~(\ref{eq:cross-corr-qpc}) is a signature of channels supporting
Majorana fermions. Importantly, this absence occurs while the QPC is
proven to actually scatter the fermions because of the dependence on
$R$ and $T$.

Our results for the zero-temperature conductance and noise properties
of normal electron and Majorana interferometers in a two-terminal
(Mach-Zehnder) and four-terminal (Hanbury Brown-Twiss) setup are
summed up in Table \ref{tab:1}.
\begin{table}[h]
  \centering
  \begin{tabular}[c]{r||c|c}
  \hline\hline
    & Normal & Majorana
		\\\hline\hline
    $G_{12}^{\mathrm{MZ}}[e^2/h]$ & $\frac 12[1+\cos(2\pi\phi/\phi_0)]$ &
		$\cos(2\pi\phi/\phi_0)$
		\\\hline
		$S_{22}^{\mathrm{MZ}}[e^3 V/h]$ & 
                $ \frac 18 [ 1 - \cos(4\pi \phi/\phi_0)]$ &
		$\frac 12 [1-\cos(4\pi\phi/\phi_0)]$
		\\\hline
    $G_{12}^{\mathrm{HBT}}[e^2/h]$ &   1/4 & 0
		\\\hline
		$S_{22}^{\mathrm{HBT}}[e^3V/h]$ &  $1/4$  &   $1$
		\\\hline
    $S_{24}^{\mathrm{HBT}}[e^3V/h]$ &
    $-\frac{1}{4}[1+\cos(2\pi\phi/\phi_0)]$ &
    $-\cos(2\pi\phi/\phi_0)$ \\
    \hline\hline
  \end{tabular}
  \caption{Summary of conductance and noise properties of normal
    electron (as in Ref.~\onlinecite{Samuelsson2004} for the HBT
    setup) and Majorana interferometers at zero temperature. In the
    Mach-Zehnder (MZ) interferometer, 1(2) labels the incoming
    (outgoing) lead. In the Hanbury Brown-Twiss (HBT) interferometer, 1
    and 3 (2 and 4) refer to the incoming (outgoing) leads.}
  \label{tab:1}
\end{table}

In conclusion, we have analyzed a Hanbury Brown-Twiss type
interferometer for Majorana fermions. We have calculated its
conductance and noise properties.  The sign of the cross-correlations
of the outgoing currents of the interferometer is predicted to be
positive if the parity of the number of vortices threading the
superconductor is odd. Our main results are three signatures for the
Majorana nature of the transport channels defined by domain walls
between superconducting and magnetic regions placed on the surface of
a three-dimensional topological insulator. On the one hand, the
average charge current in the outgoing leads vanishes since there are
symmetric probabilities for outgoing electrons or holes, see the
discussion before Eq.~(\ref{two-particle-inf}).  This vanishing
conductance needs to be complemented by a check that the structure is
functional, which is provided by the finite current auto-correlation.
On the other hand, we find a finite zero-temperature shot noise at the
output port of the interferometer even for a vanishing average current
reflecting the finite fluctuations of the Majorana particle around
charge neutrality. Finally, our calculations predict the absence of
electronic partition noise in a quantum point contact, whereas the
parameter dependence of the scattering matrix proves that the point
contact actually scatters the fermions. These signatures will be an
important help in verifying the existence of Majorana excitations in
interferometric structures at the surface of topological insulators.

\acknowledgments 
GS and CB acknowledge financial support by the EC IST-FET
project SOLID, the Swiss SNF, the NCCR Nanoscience, and the NCCR
Quantum Science and Technology.
MSC has been supported by the NRF Grants (2009-0080453, 2010-0025880,
2011-0012494). WB was financially supported by the DFG through SFB 767
and SP 1285.

\end{document}